\documentclass{ws-procs9x6-cpt16}
\begin{document}

\newcommand{\refeq}[1]{(\ref{#1})}
\def\etal {{\it et al.}}

\def\al{\alpha}
\def\be{\beta}
\def\ga{\gamma}
\def\de{\delta}
\def\ep{\epsilon}
\def\ve{\varepsilon}
\def\ze{\zeta}
\def\et{\eta}
\def\th{\theta}
\def\vt{\vartheta}
\def\io{\iota}
\def\ka{\kappa}
\def\la{\lambda}
\def\vpi{\varpi}
\def\rh{\rho}
\def\vr{\varrho}
\def\si{\sigma}
\def\vs{\varsigma}
\def\ta{\tau}
\def\up{\upsilon}
\def\ph{\phi}
\def\vp{\varphi}
\def\ch{\chi}
\def\ps{\psi}
\def\om{\omega}
\def\Ga{\Gamma}
\def\De{\Delta}
\def\Th{\Theta}
\def\La{\Lambda}
\def\Si{\Sigma}
\def\Up{\Upsilon}
\def\Ph{\Phi}
\def\Ps{\Psi}
\def\Om{\Omega}
\def\cA{{\cal A}}
\def\cB{{\cal B}}
\def\cC{{\cal C}}
\def\cE{{\cal E}}
\def\cF{{\cal F}}
\def\cG{{\cal G}}
\def\cL{{\cal L}}
\def\cM{{\cal M}}
\def\cO{{\cal O}}
\def\cP{{\cal P}}
\def\cR{{\cal R}}
\def\cV{{\cal V}}
\def\fr#1#2{{{#1} \over {#2}}}
\def\half{{\textstyle{1\over 2}}}
\def\quar{{\textstyle{1\over 4}}}
\def\eigh{{\textstyle{1\over 8}}}
\def\frac#1#2{{\textstyle{{#1}\over {#2}}}}
\def\vev#1{\langle {#1}\rangle}
\def\expect#1{\langle{#1}\rangle}
\def\norm#1{\left\|{#1}\right\|}
\def\abs#1{\left|{#1}\right|}
\def\lsim{\mathrel{\rlap{\lower4pt\hbox{\hskip1pt$\sim$}}
    \raise1pt\hbox{$<$}}}
\def\gsim{\mathrel{\rlap{\lower4pt\hbox{\hskip1pt$\sim$}}
    \raise1pt\hbox{$>$}}}
\def\sqr#1#2{{\vcenter{\vbox{\hrule height.#2pt
         \hbox{\vrule width.#2pt height#1pt \kern#1pt
         \vrule width.#2pt}
         \hrule height.#2pt}}}}
\def\prt{\partial}

\def\mn{{\mu\nu}}
\def\rs{{\rh\si}}
\def\kl{{\ka\la}}
\def\b{q}
\def\bHat{\widehat{\b}}
\def\c{s}
\def\cHat{\widehat{\c}}
\def\cd#1{{\c}^{(#1)}}
\def\d{k}
\def\dHat{\widehat{\d}}
\def\mbf#1{\mbox{\boldmath$#1$}}
\def\pvec{\mbf p}
\def\phat{\mbf{\hat p}}
\def\syjm#1#2{\phantom{}_{#1}Y_{#2}}
\def\kjm#1#2#3{k^{(#1)}_{(#2)#3}}
\def\kI{\kjm{d}{I}{jm}}
\def\kE{\kjm{d}{E}{jm}}
\def\kB{\kjm{d}{B}{jm}}
\def\kV{\kjm{d}{V}{jm}}
\def\kVt{\kjm{3}{V}{jm}}
\def\kIdjm#1#2{\kjm{#1}{I}{#2}}
\def\kEdjm#1#2{\kjm{#1}{E}{#2}}
\def\kBdjm#1#2{\kjm{#1}{B}{#2}}
\def\kVdjm#1#2{\kjm{#1}{V}{#2}}

\title{Nonminimal Lorentz violation}

\author{Matthew Mewes}

\address{Physics Department, California Polytechnic State University\\
San Luis Obispo, CA 93407, USA}

\begin{abstract}
This contribution to the CPT'16 meeting provides
a brief overview of recent studies
of nonminimal Lorentz violation
in the Standard-Model Extension.
\end{abstract}

\bodymatter

\phantom{}\vskip10pt\noindent
The Standard-Model Extension (SME) provides
a general field-theoretic description
of violations of Lorentz and CPT invariance
in particle physics and in gravity
and has facilitated hundreds of experimental tests
of these fundamental symmetries.\cite{datatables}
A Lorentz-violating term in
the lagrangian of the SME
is constructed from a tensor coefficient
contracted with a conventional tensor operator,
giving contributions to the action
that schematically take the form
\begin{equation}
\de S = \int d^4x\
({\rm coefficient~tensor})
\cdot({\rm tensor~operator}) \ .
\end{equation}
The coefficients for Lorentz violation
give the vacuum a non-scalar structure
and act as Lorentz-violating background fields.
The various violations are often classified
according to the mass dimension $d$
of the associated operator.

Early development of the SME
focused largely on minimal violations
--- those involving Lorentz-violating operators
with renormalizable mass dimensions $d=3$ and $4$ ---
leading to the so-called minimal SME (mSME).
Pioneering works constructed
the minimal modifications
to the Standard Model of particle physics\cite{ck}
followed by General Relativity.\cite{sme:grav}
While the mSME has served as the theoretical basis
for a vast majority of Lorentz tests
performed to date,\cite{datatables}
more recent efforts aim at exploring
nonminimal violations associated
with operators of mass dimensions $d\geq 5$.
These include the construction of
the full nonminimal extensions for
free photons,\cite{km:ph}
neutrinos,\cite{km:nu}
free Dirac fermions,\cite{km:ferm}
and linearized gravity.\cite{bkx:grav:sr,kt:grav:cr,km:grav}

The study of Lorentz-violating operators of
nonrenormalizable dimensions can be motivated
through simple dimensional analysis.
For operators of dimension $d$,
the coefficients for Lorentz violation
have mass dimension $4-d$.
Assuming a connection to Planck-scale physics,
we might naively expect coefficients that
are of order $\sim M_P^{4-d}$,
where $M_P$ is the Planck mass.
This type of suppression gives
the SME the structure
of a series approximation,
which fits well with the notion
that it represents the low-energy limit
of some underlying high-energy theory.
The idea is that conventional Lorentz-invariant physics
(the Standard Model and General Relativity)
corresponds to the dominant leading-order terms in the series.
The Lorentz-violating terms in the SME
give higher-order corrections,
which decrease in size with higher dimension $d$.
This makes sense for nonminimal violations with $d\geq 5$,
but is problematic for minimal violations,
where $d=4$ violations would be unsuppressed
and $d=3$ violations would be very large.
Minimal violations would be easily detected in this picture
and precluded by their absence.
The lowest-order Lorentz violations
would then be nonminimal ones
with mass dimension $d\geq 5$.

While the mSME has been tested extensively,
the nonminimal parts of the SME
remain relatively unexplored,
with most of the parameter space unconstrained
and completely open to future experimentation.
At present, constraints on
nonminimal Lorentz violation in photons
include tests involving
vacuum birefringence,\cite{km:ph,stecker:bire}
vacuum dispersion,\cite{disp}
and resonant cavities.\cite{cav}
In neutrinos, there are constraints from
oscillation experiments\cite{km:nu}
and kinematical tests involving
high-energy astrophysical neutrinos.\cite{km:nu,nu:astro}
For heavier fermions,
current constraints on nonminimal violations
come from studies of muonic atoms\cite{akv:muons}
and hydrogen-like atoms.\cite{kv:hyd}
Bounds on nonminimal Lorentz violation in gravity
have been obtained
in tests of short-range gravity,\cite{bkx:grav:sr,sr:grav:exp}
from limits on gravitational \v Cerenkov radiation
in cosmic rays,\cite{kt:grav:cr}
and from limits on birefringence
in gravitational waves.\cite{km:grav}

The different sectors of the nonminimal SME
share some common features,
which we illustrate here
by considering the recently constructed
nonminimal extension for linearized General Relativity.\cite{km:grav}
The lagrangian for this extension,
in term of the metric perturbation $h_\mn$,
can be written
\begin{eqnarray}
\cL &=&
\tfrac14 \ep^{\mu\rh\al\ka} \ep^{\nu\si\be\la} 
\et_\kl h_\mn \prt_\al\prt_\be h_\rs
\nonumber\\&&
+\tfrac14 h_\mn \cHat^{\mn\rs} h_\rs
+\tfrac14 h_\mn \bHat^{\mn\rs} h_\rs
+\tfrac14 h_\mn \dHat^{\mn\rs} h_\rs\ ,
\label{lag}
\end{eqnarray}
where the first term is the usual Lorentz-invariant part.
The term involving the tensor operator
$\cHat^{\mn\rs}$ contains CPT-even Lorentz violations for even $d\geq4$,
$\bHat^{\mn\rs}$ gives CPT-odd violations with odd $d\geq 5$,
and $\dHat^{\mn\rs}$ gives CPT-even violations with even $d\geq6$.
Each of the tensor operators represents
an infinite series of derivatives.
For example,
$\cHat{}^{\mu\rh\nu\si}$ is given by the expansion
$\cHat{}^{\mu\rh\nu\si} =
\sum_d \cd{d}{}^{\mu\rh\ep_1\nu\si\ep_2\ep_3\ldots\ep_{d-2}}
\prt_{\ep_1}\prt_{\ep_2}\prt_{\ep_3}\ldots\prt_{\ep_{d-2}}$.
The expansion coefficients
$\cd{d}{}^{\mu\rh\ep_1\nu\si\ep_2\ep_3\ldots\ep_{d-2}}$
are constant and control the Lorentz violation.
The differing symmetries of the coefficients in each class
result in different physical consequences.

The minimal version of the above theory is obtained
by restricting attention to $d=4$ $\cHat^{\mn\rs}$ terms,
which has the effect of demoting the operator
$\cHat^{\mn\rs}$ to the constant $\c^{(4)\mn\rs}$.
In other sectors of the SME,
going the other way,
from the minimal to the nonminimal extension,
is effectively the reverse of this procedure.
In almost every case,
the introduction of nonminimal violations
can be viewed as the promotion of minimal $d=3,4$ coefficients
from constants to operators.\cite{km:ph,km:nu,km:ferm}
The exception is linearized General Relativity,
where the operators $\bHat^{\mn\rs}$ and $\dHat^{\mn\rs}$
lack minimal counterparts.

The power of the SME lies in its ability to
give theoretically consistent predictions for
almost any system.
For example,
with the theory in Eq.\ \refeq{lag}
we can examine the effects
of general Lorentz violation on gravitational waves.
A key result is the modified dispersion relation,
$\om = \big(
1-\vs^0 \pm \sqrt{(\vs^1)^2+(\vs^2)^2+(\vs^3)^2}\,\big)|\pvec|$,
which predicts several unconventional features,
including dispersion and birefringence.
The $\vs^a$ parameters are complicated momentum-dependent
combinations of the coefficients for Lorentz violation.
However, as in many other practical applications of the SME,
the problem can be made tractable by performing
a spherical-harmonic decomposition.
This aids in the cataloging of
the many Lorentz-violating effects that can arise.
It also significantly simplifies the rotations
between noninertial laboratory frames
and the inertial Sun-centered frame
conventionally used for reporting results.\cite{sunframe}

In the current example,
the spherical expansion of the $\vs^a$ parameters is
\begin{eqnarray}
\vs^0 &=& 
\sum \om^{d-4} \, \syjm{}{jm}(-\phat)\,  \kI\ ,
\quad
\vs^3 =
\sum_{djm} \om^{d-4} \, \syjm{}{jm}(-\phat)\, \kV\ ,
\nonumber \\
&& \quad
\vs^1\mp i \vs^2 =
\sum \om^{d-4} \, \syjm{\pm4}{jm}(-\phat)\,
\big(\kE\pm i\kB\big)\ .
\end{eqnarray}
The spherical coefficients for Lorentz violation
$\kI$, $\kE$, $\kB$, and $\kV$
completely characterize the leading-order effects
of Lorentz violation in gravitational waves.
Currently,
the $\kI$ coefficients have been bounded by
the apparent absence of
gravitational \v Cerenkov radiation
in cosmic rays.\cite{kt:grav:cr}
The $\kE$, $\kB$, and $\kV$ coefficients
are constrained by limits on birefringence
in gravitational waves.\cite{km:grav}

Note that the above expansion involves
spin-weighted spherical harmonics $\syjm{s}{jm}$.
These are similar to the familiar harmonics $Y_{jm}=\syjm{0}{jm}$,
but carry spin weight $s$,
which is equivalent to helicity,
up to a sign.
The appearance of the $s=\pm 4$ harmonics
in the above example stems from
the spin-2 nature of gravitational waves.
The violations associated with the $s=\pm4$ harmonics
couple the $\pm 2$-helicity waves,
leading to changes of $\pm 4$ in helicity.
Similarly,
the $s=\pm2$ harmonics arise
for spin-1 photons in the SME,\cite{km:ph}
and $s=\pm1$ harmonics appear in
the spherical expansions
for spin-$\half$ fermions.\cite{km:nu,km:ferm}

\section*{Acknowledgments}
This work is supported in part by
the United States National Science Foundation
under grant PHY-1520570.

\end{document}